\documentclass{appolb}
\usepackage{
	epsfig
	,amsmath
	,graphicx
}

\begin{document}
\title{Accessibility of color superconducting quark matter phases 
in heavy-ion collisions
}
\author{
	D.~B.~Blaschke
\address{Institute for Theoretical Physics,
	University of Wroc{\l}aw, 50-204 Wroc{\l}aw, Poland\\
	Bogoliubov Laboratory of Theoretical Physics, JINR, 
        141980 Dubna, Russia
}
\and
	F.~Sandin
\address{
	IFPA, D\'epartement AGO, Universit\`e de Li\`ege, Sart Tilman, 
4000 Li\`ege, Belgium\\
EISLAB, Lule{\aa} University of Technology, 971 87 Lule{\aa}, Sweden
}
\and
	V.~V.~Skokov and S.~Typel
\address{
	GSI Helmholtzzentrum f\"ur Schwerionenforschung GmbH, Theorie,\\ 
D-64291 Darmstadt, Germany
}
}
\maketitle
\begin{abstract}
We discuss a hybrid equation of state (EoS) that fulfills constraints
for  mass-radius relationships and cooling of compact stars. 
The quark matter EoS is obtained from a 
Polyakov-loop Nambu--Jona-Lasinio 
(PNJL) model with color superconductivity, and the hadronic one from a 
relativistic mean-field (RMF) model with density-dependent couplings (DD-RMF).
For the construction of the phase transition regions we employ here for
simplicity a Maxwell construction.
We present the phase diagram for symmetric matter which exhibits two 
remarkable features: (1) a ``nose''-like structure of the hadronic-to-quark 
matter phase border with an increase of the critical density at 
temperatures below $T \sim 150$ MeV and (2) a high critical temperature for 
the border of the two-flavor color superconducting (2SC) phase, $T_c > 160$ 
MeV.
We show the trajectories of heavy-ion collisions in the plane of excitation
energy vs. baryon density calculated using the UrQMD code 
and conjecture that for incident energies of $4 \dots 8$ A GeV as provided,
e.g., by the Nuclotron-M at JINR Dubna or by lowest energies at the future 
heavy-ion collision experiments CBM@FAIR and NICA@JINR, 
the color superconducting quark matter phase becomes accessible.

\end{abstract}
\PACS{11.10.Wx, 11.30.Rd, 12.38.Mh}
  
Theoretical studies of the QCD phase diagram have predicted a rich 
structure of nonperturbative phases under conditions of   
temperatures $T$ below the deconfinement temperature $T_c \sim 180$ MeV 
found in lattice QCD studies \cite{Bazavov:2009zn} and baryochemical 
potentials $\mu_B$  above $\sim m_N$, where $m_N=939$ MeV is the nucleon mass. 
Of particular interest are the questions: 
\begin{itemize}
\item How does the order and the location of the chiral phase transition 
depend on temperature, density, size and isospin asymmetry of the system? 
\item What is the nature of confinement and how does deconfinement occur?
\item Can deconfinement and chiral symmetry restoration 
occur independent from each other at high densities?  
As a consequence, shall we expect massive deconfined quark matter or chirally 
symmetric but confined quark matter (quarkyonic matter)?
\item Is dense quark matter (color) superconducting? Does confinement preclude 
color superconductivity? Is there a BEC or rather BCS phase of color 
superconductivity? What is the critical temperature?
Can these phases be created in the laboratory?
\end{itemize}
The energy scan program of the NA49 experiment has given indications for
a phase change at $E\sim 30$ A GeV, in particular from the peak (``horn'') in
the $K^+/\pi^+$ ratio. 
Recently, it has been suggested that the ``horn'' may be the signature of an
approximate triple point in the QCD phase diagram \cite{Andronic:2009gj}
where three phases meet:  hadronic matter, quarkyonic matter, and a 
quark-gluon plasma.
Experiments of the next generation (NA61-SHINE, 
low-energy RHIC, CBM and NICA) should, however, take into their focus the 
possibility that qualitatively new features could be found at still lower 
energies.
This concerns in particular color superconducting quark matter phases like 
the 2SC phase \cite{Zablocki:2009ds}
and the conjectured quarkyonic phase \cite{McLerran:2007qj}.
At the JINR Dubna, the modernized nuclotron-M and the planned nuclotron-based 
ion collider facility (NICA) give a unique opportunity to explore the above 
mentioned region of the phase diagram, and may thus complement alternative 
programs for systematic studies of heavy-ion collisons in the relevant range 
of collision energies $2\le E\le 40$ A GeV.

\begin{figure}[thb]
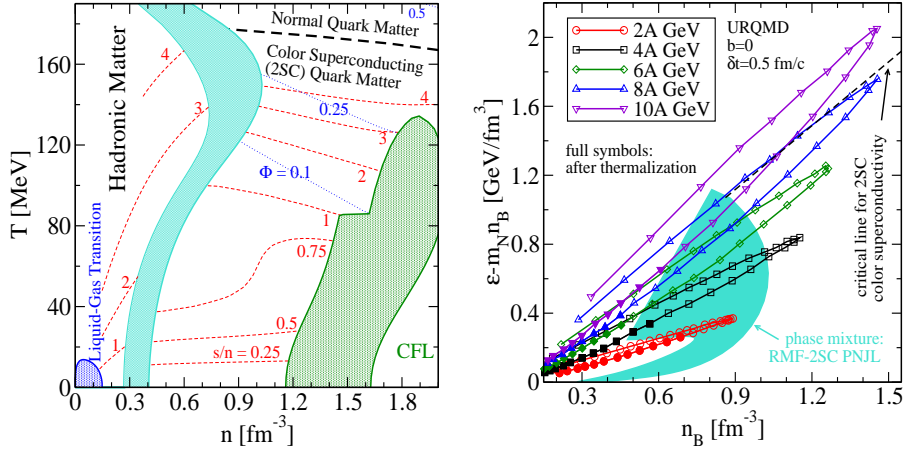

\begin{tabular}{cc}
\includegraphics[height=0.47\textwidth,width=0.45\textwidth]{NICA-PHD.eps}
&
\includegraphics[height=0.47\textwidth,width=0.45\textwidth]{Tra-Tra_URQMD.eps}
\end{tabular}
\caption{Left panel: QCD model phase diagram with mixed phase regions 
corresponding to the first order phase transitions: nuclear liquid-gas (blue), 
hadron - quark matter (turqoise), 2SC - CFL quark matter (green).
The transition from color superconducting (2SC) quark matter to normal quark 
matter is of second order (dashed line).
Right panel: trajectories of contral (b=0) heavy-ion collisions at different 
energies in the excitation energy-density plane overlayed to the hadronic 
matter - 2SC quark matter mixed phase region of the model-QCD phase diagram.
The hatched region indicates the mixed phase.
The dashed line denotes the critical line for 2SC color superconductivity. 
}
\end{figure}

As it has been demonstrated in \cite{Roessner:2006xn,GomezDumm:2008sk} the 
coupling to the Polyakov loop increases the critical temperature for the 
2SC phase to the order of the deconfinement temperature $T_{\rm 2SC}\sim 150$
MeV, see the left panel of Fig.~1.
In that figure, we show a modern QCD model phase diagram based on
a quark matter EoS from a three-flavor NJL model with selfconsistent quark 
masses and diquark gaps 
\cite{Blaschke:2005uj,Ruester:2005jc,Warringa:2005jh,Abuki:2005ms}, 
generalized here by the coupling to the Polyakov-loop potential 
to suppress unpysical quark degrees of freedom. 
The hadronic phase is modeled with a density-dependent relativistic
meanfield approach \cite{Typel:2005ba} which also describes the nuclear 
liquid-gas phase transition with a critical point, see the blue hatched 
region in Fig.~1 (left panel). The hadron-to-quark matter transition is 
obtained from a Maxwell construction with a mixed phase coexistence region 
shown by the turqoise hatched region.
The unusual nose-like shape of this region is due to the Polyakov-loop 
potential which suppresses the quark pressure at finite temperatures below the 
deconfinement temperature, but not at $T=0$. At low temperatures, the 
appearance of the diquark condensate shifts the chiral restoration transition
to rather low densities, of the order of $2-3~n_0$, $n_0=0.16$ fm$^{-3}$.

In order to answer the question of the accessibility of these novel phases of
dense QCD matter, we have examined the trajectories of the Lorentz contracted 
central region of central $Au-Au$ collisions of given energies in the range 
$2<E<10$ AGeV from UrQMD simulations, see the right panel of Fig.~1. 
The hatched region corresponds to the mixed phase of hadronic and 2SC 
quarkyonic matter for the parametrization of the PNJL model without 
vector mean field. 
The dashed line denotes the critical line for 2SC color superconductivity. 
We conclude from this figure that for energies $4<E<8$ AGeV, which are 
accessible by the present nuclotron-M facility, one may expect to enter the 
2SC color superconducting quark matter phase with restored chiral symmetry
and strong color correlations due to a low Polyakov-loop meanfield $\Phi<0.25$,
indicating a quarkyonic phase \cite{Fukushima:2008wg}.
The exploration of the transition from color superconducting to normal quark 
matter and finally the ceasing of the mixed phase at the QCD critical point 
would require energies beyond 10 AGeV, aimed to be reached at NICA.

Finally, let us discuss two ideas for the experimental identification of the 
chiral restoration and the color superconductivity transition which should be
considered when planning experiments and in particular when designing the 
multi-purpose detector (MPD) system.
\begin{enumerate}
\item An enhancement of the two-photon invariant mass spectrum in the mass 
range $M_{2\gamma}\sim 300$ MeV, from the decay of the sigma meson which should
become a long-lived ``sharp'' resonance when chiral symmetry gets restored and
the dominant two-pion decay channel gets closed 
\cite{Volkov:1997dx,Chiku:1997va}. 
This signal shall also prevail in the hypothetic quarkyonic phase and 
more traditional estimates of the two-photon spectrum within the ordinary NJL
model would have to be revised within the PNJL model.
\item An enhancement of the lepton-pair invariant mass spectrum when 
approaching the critical temperature for color superconductivity from above 
(precursor effect \cite{Kunihiro:2007bx}) which should eventually turn into a 
resonance-like structure when entering the 2SC phase, due to additional 
contributions to the diquark-antidiquark annihilation diagrams 
(generalized Aslamasov-Larkin and Maki-Thompson diagrams) containing anomalous 
propagator contributions. 
\end{enumerate}
In conclusion we would like to stress that our modern QCD model phase diagram 
suggests that new dense quark matter phases (color superconductor and 
quarkyonic matter) are accessible already at the present nuclotron-M energies 
and that both the study of the transition to normal quark matter and the 
vanishing of the mixed phase at the QCD critical endpoint will require higher 
energies than presently available at the nuclotron-M but are attainable in the 
planned FAIR-CBM and NICA-MPD experiments.

\subsection*{Acknowledgements}

This work has been supported in part by the Polish Ministry of Science and 
Higher Education  (MNiSW) under grant No. N N 202 2318 37 (DB), by 
the Russian Fund for Basic research (RFBR) under grant No. 08-02-01003-a 
(DB,VVS), by FNRS, the Belgian fund for scientific research (FS), by the DFG 
cluster of excellence ``Origin and Structure of the Universe'' (ST) and by 
CompStar, a research networking programme of the European Science Foundation.

\end{document}